\documentclass[10pt, conference, compsocconf]{IEEEtran}
\usepackage{amssymb}
\usepackage{cite}
\usepackage[cmex10]{amsmath}
\usepackage{algorithm,algpseudocode}
\usepackage{array}
\usepackage{url}
\usepackage[usenames,dvipsnames]{color}

\ifCLASSINFOpdf
  \usepackage[pdftex]{graphicx}
  \graphicspath{{../pdf/}{images/}}
  \DeclareGraphicsExtensions{.pdf,.jpeg}
\else
  \usepackage[dvips]{graphicx}
  \graphicspath{{../eps/}}
  \DeclareGraphicsExtensions{.eps}
\fi

\ifCLASSOPTIONcompsoc
  \usepackage[caption=false,font=normalsize,labelfont=sf,textfont=sf]{subfig}
\else
  \usepackage[caption=false,font=footnotesize]{subfig}
\fi

\newtheorem {theorem}{Theorem}

\newtheorem {lemma}[theorem]{Lemma}

\renewcommand{\baselinestretch}{0.99}

\newcommand{\eat}[1]{}

\newcommand{\hidenote}[1]{}

\begin{document}
\title{How to Return to Normalcy: Fast and Comprehensive Contact Tracing of COVID-19 through Proximity Sensing Using Mobile Devices}

\author{
\IEEEauthorblockN{Ye Xia\\}
\IEEEauthorblockA{Department of Computer and Information Science and Engineering\\
University of Florida, Gainesville, FL 32611\\}
Email: yx1@ufl.edu
\and
\IEEEauthorblockN{Gwendolyn Lee\\}
\IEEEauthorblockA{Warrington College of Business \\
University of Florida, Gainesville, FL 32611\\
Email: gwendolyn.lee@warrington.ufl.edu }
}


\maketitle

\begin{abstract}
We outline a contact-tracing strategy based on proximity sensing using mobile devices. We discuss what an ideal system should look like and what it can do. We show that, when adopted sufficiently broadly, such a contact-tracing strategy can bring COVID-19 under complete control, end the need of social distancing, and return the society to full normalcy. We also review some of the challenges faced by the current generation of proximity-sensing technologies, including Bluetooth Low Energy used by phones, and consider both interim and longer-term solutions. Our main contribution is that we reason through why such a contact-tracing strategy is likely to achieve the stated goal of returning to full normalcy. Using probabilistic models, we show that universal adoption is not necessary to achieve the stated goal, thus there is some room for exceptions; however, the adoption rate needs to be very high, e.g., above $95\%$ depending on the disease parameters. With more vigilance in disease surveillance to detect mild cases earlier, the number may be brought down to about $90\%$. The results call for deployment effort to be led by public authorities at the state or federal level so that the required adoption rate can be reached and the tracing coverage is wide enough to be relevant for disease control.
\end{abstract}


\section{Introduction} \label{sec:intro}

The motivating question we ask here is what comes next after the strong suppression measures of COVID-19 succeed. Imagine that the suppression measures have worked so well in a country or a major city that only a small number of cases are reported daily. Can people return to the kind of life and work they had before the outbreak of COVID-19? Unless herd immunity has been achieved or a curative treatment has been discovered, the answer is likely no because a few cases a day will start a new cycle of exponential increase, turning into thousands of cases a day in a matter of weeks or months.  The current proposals for economic reopening all include ubiquitous social distancing. As long as social distancing is in place, there is no normalcy. To return to genuine normalcy and sustain it, people must feel the risk of infection is low and it must remain that way indefinitely. In other words, the state of having a low number of infections in the population must be a {\em steady} one rather than merely the beginning stage of an exponential growth.

To achieve that steady state, we envision a technology-based contact-tracing strategy that makes contact tracing more rapid, more accurate, more comprehensive and less labor-intensive. We will argue that if it is adopted sufficiently widely and if the infected individuals discovered through tracing are quarantined (which requires widespread testing and solid quarantine measures), the number of infections can be kept very low indefinitely, allowing the society to return to its normal functioning.

The key of this contact-tracing strategy is a mobile device that all or a majority of the people shall carry most of the time when they go to work, shops, or other public places. Let's call this device a {\em contact recorder}. Its main function is to detect the presence of other individuals who also carry contact recorders in the vicinity, say within 2 meters, and record those device IDs, the distance, and the duration of contact at the distance. Once an infected person, say $A$, is identified by testing, the recorded contact information is used to find all the people that $A$ might have infected through close contact. The contact information can also help to identify who might have infected $A$, even if that person is asymptomatic.

To do that, the contact recorder needs to be equipped with hardware capable of short-range distance sensing, also known as {\em proximity sensing}. Most of the newer phones have a form of such proximity sensing capability based on Bluetooth Low Energy (BLE) -- by measuring the received signal strength. The required proximity sensing is not something the GPS system can alone provide. Although the GPS trajectory data have some usefulness for the intended application (see later), they are not accurate enough for detection of short-range contacts. 

It is important to emphasize that the benefit of the proposed strategy is not just that a person who carries the device will be notified quickly if he/she was in contact with an infected individual. Its main benefit, when deployed and used broadly enough, is to ensure that the risk of infection for everyone is low to start with, so that people don't have worries about resuming their normal life. 

In this paper, we outline the above described contact-tracing strategy. We discuss what an  ideal system should look like and what it can do. We also review some of the challenges faced by the current generation of proximity-sensing technologies. In the interest of timely deployment to cope with the current crisis, the contact recorder can be initially implemented as an app on a smart watch or a mobile phone. In the long run, the challenges should be dealt with more carefully and the device may need to be designed from ground up in anticipation of the next pandemic. We also speculate why such a contact-tracing strategy can be accepted widely, wider than what first impression may suggest.

One of our main contributions is that we reason through why such a contact-tracing strategy is likely to achieve the goal of returning to full normalcy. One of our main messages is that the potential of such contact tracing is huge: It can stop the spread of the disease and keep it that way without social distancing. Another main message is that universal adoption is not necessary to achieve the stated goal, thus there is some room for exceptions; however, the adoption rate needs to be very high, e.g., above $95\%$ depending on the disease parameters. With more vigilance in disease surveillance to detect mild cases earlier, the number may be brought down to about $90\%$. The results call for deployment effort to be led by public authorities at the state or federal level so that the required adoption rate can be reached and the tracing coverage is wide enough to be relevant for disease control.

As of April 2020, we are seeing an increasing number of proposed or actual contact-tracing systems based on proximity sensing \cite{TraceTogether, AppleGoogle, DP3T, PEPP}. Singapore has deployed a phone-based system, called {\em TraceTogether} \cite{TraceTogether}, which contains a subset of the features outlined in this paper. The contact data of an individual is stored in his/her phone. Once the individual is confirmed to be infected, he/she can choose to allow the government server to access the data in the phone app, which helps identify close contacts. Apple and Google have made announcement that they will launch a comprehensive solution at the level of application programming interfaces (APIs) and operating systems to assist in enabling contact tracing using apps developed by third-party developers and/or public health authorities \cite{AppleGoogle}. User privacy features strongly in their solution. For instance, explicit user consent is required to use the contact data, personally identifiable information or user location data will not be collected, and the list of contacted people never leaves a user's phone. European research institutions and governments are behind the proposals in \cite{DP3T, PEPP}, which also have strong privacy-protection components. In particular, \cite{DP3T} takes a decentralized approach to privacy-preserving tracing.

We feel that these proposals have not focused enough on the full potential of such contact-tracing systems. They exhibit a somewhat laissez-faire philosophy for adoption and use of such systems. First, for perfectly understandable reasons of not looking draconian, the advocacy for adoption is drastically toned down with messages such as `people should pass words around and opt in'. Second, the contact information is only made available incrementally. When tracing reaches an individual, a request is sent and contact information becomes available after the individual's consent \cite{TraceTogether}. We make stronger claims here and we attempt to substantiate them. The contact-tracing strategy based on proximity sensing can keep the disease fully contained without the need of other mitigation measures such as social distancing or universal mask-wearing. It is not just a nice thing to have that complements other measures; it can end social distancing and allow the society to return to full normalcy. However, in order to achieve that, the adoption rate needs to be very high. We show that the ability to trace an entire cluster of infections is important, rather than just a subset of the individuals on best-effort basis. For that, it is crucial that the contact information from all individuals in the cluster is available, and preferably in real time to minimize the response time. Other proposals that use more laid-back approaches do not allow easy tracing of an entire cluster.

The remainder of the paper is organized as follows. In Section \ref{sec:mainidea}, we describe the main idea of the contact-tracing strategy. In Section \ref{sec:whyeffective}, we answer why the strategy will achieve the stated goal. A highlight is that we analyze the minimum adoption rate required for the strategy to work as intended. In Section \ref{sec:interimlongterm}, we discuss the device requirements, and the interim and long-term solutions. The conclusion and additional discussions are given in Section \ref{sec:conclude}.

\section{The Main Idea of the Strategy}
\label{sec:mainidea}

\subsection{Why Fast and Comprehensive Contact Tracing is Crucial} \label{sec:wf_model}

The experiences of South Korea, Singapore, Hong Kong, and Taiwan have provided evidence that it is possible to keep the number of COVID-19 cases at a low level for a long period of time through aggressive testing, relentless contact tracing and high-quality quarantine. The reason these measures work is that together they can reduce the {\em reproduction ratio}, which is the average number of individuals directly infected by one patient. The reproduction ratio is a key parameter governing the dynamics of the number of infected cases. When the reproduction ratio is greater than $1$, with a very high chance, the number of infections grows exponentially fast. Clearly, the smaller the reproduction ratio, the slower the number of infections grows. If the ratio is reduced to below 1, the exponential growth will not happen and the number of infections will collapse until it reaches zero. For COVID-19, earlier estimates put the basic reproduction ratio, usually denoted by $R_o$, in the range between $3$ to $4$. One recent estimate put it as high as $5.7$ \cite{CDC20}. It is not clear whether the measures taken by aforementioned Asian regions have reduced it to a number below $1$ or just slightly above $1$.

In a perfect world where the three measures work perfectly and without delay, any infected individual will be immediately isolated from the general population (the reproduction ratio will be zero). Realistically, they each have some challenges. Among the three measures, how to improve the testing capability and speed has attracted a great deal of effort and it is expected that testing will soon not be a bottleneck. It is also hopeful that countries and regions will eventually see the value of and institute solid quarantine processes.\footnote{If the proposed contact-tracing strategy works as intended in keeping the number of infections low, the required testing and quarantine capabilities will be more easily satisfied.} Contact tracing remains a problem, as traditional contact tracing requires boots on the ground, and therefore, is labor-intensive and suffers long delay. Importantly, it suffers a more difficult, inherent traceability challenge. A patient may be unable to recall all the contacts he/she made, and anonymous contacts in public places are not traceable. Once community-based infection becomes serious enough, societies often abandon contact tracing altogether.

We argue that the difficulties in contact tracing can be largely resolved by using technologies. In other words, contact tracing can be made more rapid, more accurate, more comprehensive and less labor-intensive. Together with widespread rapid testing and proper quarantine, nearly all infected persons can be quickly isolated from the general population. In other words, such contact tracing approximates what can be done in a perfect world, and therefore, the society can be brought back to normal.

\subsection{Tracing by Contact Recorder}

The most important piece of this contact-tracing strategy is the mobile device, the {\em contact recorder}, that everyone carries in the public. Think about a smart watch or a mobile phone. Each such device has a unique ID. On a central server, the device ID is tied to the individual's identity and how the individual can be reached, such as a cell phone number or an email address.

When two individuals are within a distance where virus transmission is possible, say 2 meters, the devices are able to detect each other and they each record the other device ID, the distance, and the duration of contact at the distance. The data is communicated to the central server, preferably in real time.

Suppose an individual, say person $A$, is sick, tested and confirmed to be infected. The central server checks the contact information recorded on $A$'s device and conducts risk evaluations of $A$'s recent contacts, say going back 14 days, and informs all those who are at risk of being infected by $A$ through contact. Depending on the risk level, some of the at-risk individuals may be tested and/or quarantined; others may only receive a notification of caution. The risk evaluation may be based on the distance of contact, the duration of contact, and how infectious person $A$ is estimated to be at the time of contact.

In addition to downstream tracing of those who might be infected by person $A$, the contact information of $A$ can also be used for upstream tracing, i.e., finding out who might have infected $A$ through contact with $A$. By repeated use of downstream and upstream tracing, a cluster of infections with the same recent origin can be identified. Another important use of upstream tracing is to identify and isolate asymptomatic or presymptomatic cases. There has been increasingly evidence that asymptomatic or presymptomatic patients can infect other people. As long as such an individual causes a serious symptomatic infection down the line, he/she can be discovered by tracing upward from the symptomatic individual.

\begin{figure}[ht]
\centering
\includegraphics[width=4in]{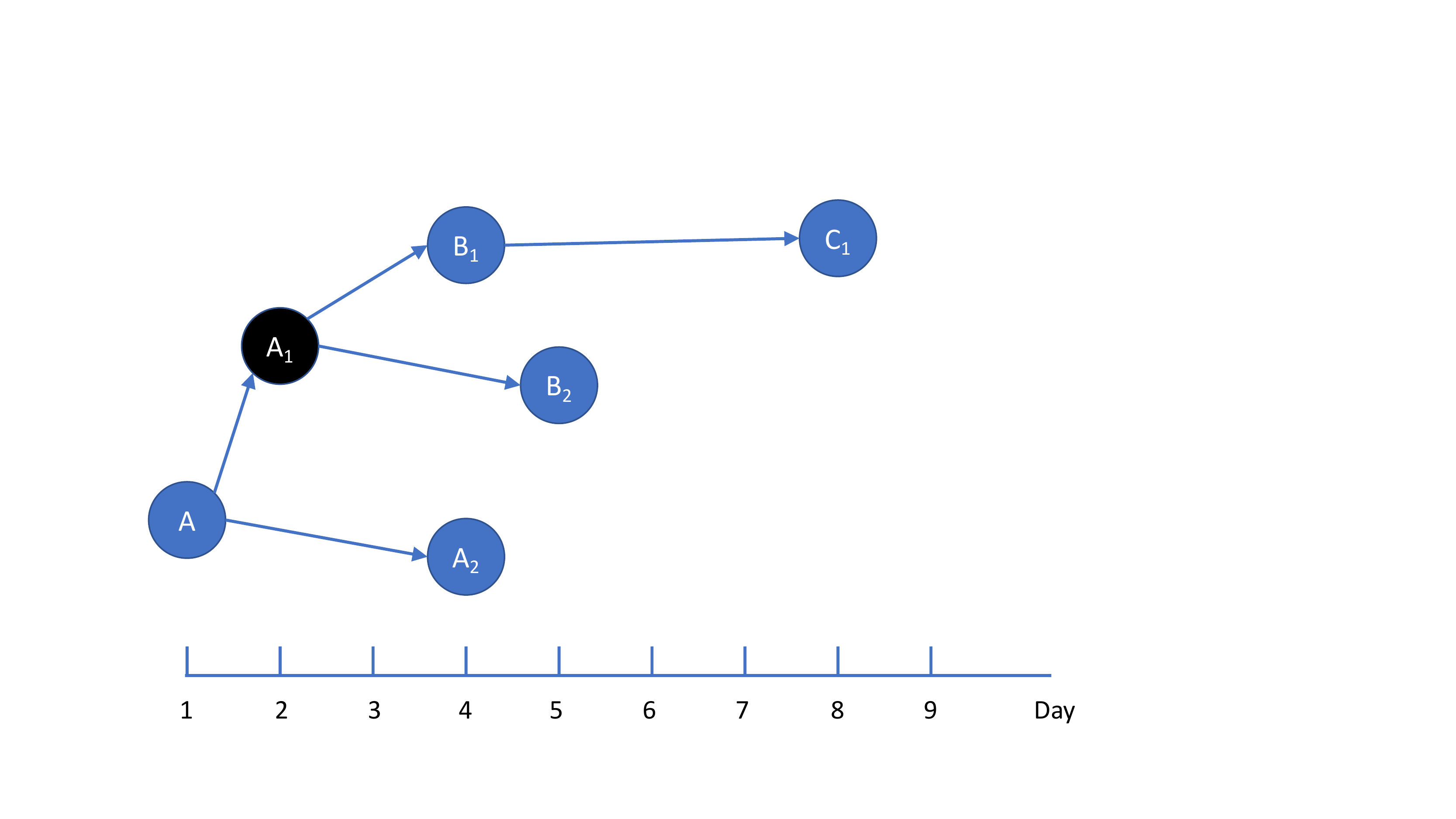}
\caption{Infection tree -- a cluster of infections started by patient $A$. $A_1$ was the first one tested and confirmed on day 9.}
\label{fig:tree}
\end{figure}

Consider an example, which is shown in Fig. \ref{fig:tree}. Suppose person $A$, who is asymptomatic, infected person $A_1$ on day 1 and $A_2$ on day 4. Suppose $A_1$ went on to infect person $B_1$ on day 4 and $B_2$ on day 5. Suppose $B_1$ infected person $C_1$ on day 8. On day 9, person $A_1$ was sick enough to seek medical attention and was tested positive. Based on $A_1$'s contact history and by testing, it is possible to confirm $A$, $B_1$ and $B_2$ as being positive. Therefore, the asymptomatic patient $A$ can be discovered. By tracing through $B_1$'s contact history, it is possible to discover $C_1$. By tracing through $A$'s contact history, it is possible to discover $A_2$. The whole cluster of patients originated from $A$ will be all accounted for. Note that the cluster can be represented as a directed graph -- in fact, a tree -- where each node represents an individual and an edge represents infection by contact. Let us call it the {\em infection tree}.

From our description of the infection events, $A$ can be recognized as being upstream to $A_1$ with respect to the path of infection. However, during actual tracing, from $A_1$'s contact history alone, it is not immediately clear whether $A$ infected $A_1$ or the other way around. More generally, the graph constructed from the contact information of all the involved individuals will initially look different from the underlying infection tree. It may have more nodes, which corresponds to contacted individuals who are not infected; the direction of the edges remain to be determined; and there may be even loops in the graph. Such a graph needs to be processed to infer the infection tree. Doing so will require not only contact information but also testing results to prune non-patients and auxiliary information such as how the disease progresses. In this paper, we will not dwell on the inference problem. When we mention a cluster of infections started by a patient $A$, we mean the infection tree rooted at $A$.

\subsection{Why Would Anyone Want to Use Such a Device?}

The primary justification for carrying a contact recorder is that it is an urgent public health measure in a crisis. There need to be promises from the authority that the measure is temporary, that its use is strictly limited to infectious disease control, and that citizens' privacy and data security are properly guarded. The use of clever privacy-preservation techniques will further alleviate worries that people may have about privacy invasion. Beyond those, there are actually incentives that may nudge citizens into using the device. The following are notable ones.

\begin{itemize}

\item	Price to pay to return to normal life: We will show in Section \ref{sec:whyeffective}, with the help of contact recorders, the number of infections can be consistently kept at a very low level so that the society can return to its normal functioning. People can go to work, restaurants, and conduct all their daily business as usual, with the comfort of knowing that the chance of infection is small. The small inconvenience of carrying the device can be viewed as the price to pay for returning to normal life.

\item	Personal benefits: The individual will benefit from getting notified very quickly should he/she be in contact with an infected person, which allows timely testing, proper quarantine (so that the individual's family is protected) and early medical care for improved outcome.

\item	Civic duty: Carrying the device can be viewed as fulfilling a civic duty required by public health measures. It can be compared to getting the required vaccines or complying with orders such as stay-at-home, social distancing, or closure of business operations or other activities. An individual who carries the device (let us call such an individual an adopter) helps everyone else while helping him/herself.

\item	Social expectation: People may be more willing to interact with an adopter because he/she can rightfully be perceived as being `less dangerous' than a non-adopter. If an adopter were in fact infectious, there would have been ample pathways in the tracing that lead to him/her. The fact that he/she is not isolated gives credence that he/she has not been infected. Nothing like that can be said about a non-adopter. Once the practice of carrying the device is adopted widely enough, it may become a social norm. 

\end{itemize}

\section{Why Will the Strategy Work?} \label{sec:whyeffective}

\subsection{Complete Traceability under Universal Adoption} \label{sec:universaladopt}

In this section, we will show that, under the assumption of universal adoption, the proposed contact-tracing strategy has the property of {\em complete traceability}, which the traditional manual method of tracing lacks. That is, assuming everyone carries a device, all consequential infections through person-to-person contact are traceable. 

We will first examine the problems of manual tracing in more details. We have already seen that, in addition to costing excessive effort and resources, there is a {\em recall problem} that afflicts manual tracing, i.e., a patient cannot recall all the contacts he/she made, especially those in public venues. There is a related problem, which we will call the {\em manifestation problem}: If an infected individual does not show any symptoms or shows only mild symptoms, he/she may not seek medical care to be tested. In principle, recall is not needed if there is no manifestation problem; every patient will show enough symptoms and will be tested. On the other hand, if there is perfect recall, the manifestation problem will go away because in a growing cluster of infections someone will show enough symptoms to be tested and everyone in the cluster will be recalled, including patients with no or mild symptoms.

In practice, the manifestation problem exists and recall is imperfect, and together, they pose severe challenges to manual tracing. For COVID-19, it is difficult to know precisely what percentage of the infected individuals are asymptomatic (or presymptomatic). Iceland tested a portion of its population and found $50\%$ of those who tested positive showed no symptoms \cite{Iceland}. Other reports vary greatly, putting the percentage as low as less than $10\%$ and as high as more than $90\%$. Much may have to do with confusing the presymptomatic cases with true asymptotic cases \cite{Diamond20}. Among the symptomatic patients, it has been estimated that between $10\% - 20\%$ of them are severe enough to require hospitalization. With the combined effects of asymptomatic, presymptomatic and mild cases, a cluster may grow to a size between one dozen to even one hundred infected individuals before one of them seeks medical attention. By then, the cluster of infected would have been in contact with a large number of people who end up not being infected, making the task of recall even more challenging. For tracing to work perfectly, the recall needs to be enough to include all the contacts that have led to infection, which is often impossible without the help of contact recorders. Consequently, manual tracing often misses many patients in a cluster, especially those asymptomatic or mild cases, as is well documented.

When tracing is done with contact recorders, both the recall and manifestation problems can be completely dealt with. We have already seen that the contact recorder solves the recall problem. With the recall problem solved, an entire cluster can be discovered as long as one patient in the cluster is tested positive. Then, what guarantees that some patient in the cluster gets tested in the first place? It is well known that, in the absence of any intervention, the cluster size will either stop growing or grow to infinity exponentially fast.\footnote{Before the population is largely immune, the time dynamic of a cluster size can be studied under the framework of branching processes. When the mean number of offsprings is greater than 1, a branching process either goes extinct or goes to infinity at an exponential rate (see page 316 in \cite{Ross96}).} If a cluster grows to infinity, it will eventually include an individual with severe enough symptoms to be caught and tested by the medical system, at which point the whole cluster will be discovered. What we mean by complete traceability is that all growing clusters are eventually fully discovered. A cluster that eventually stops growing by itself will stop contributing to the spread of the disease, and hence, can be ignored from the current discussion.

Although all growing clusters are eventually discovered with the proposed tracing strategy,  the delay in discovering a cluster and the cluster size at the time of discovery depend on many details in the tracing process, the random infection process and how the disease progresses, as shown in the example of Fig. \ref{fig:tree}. Since shorter delay and a smaller cluster size clearly have practical benefits, countries should try to make rapid testing abundantly available and the public should be informed to seek testing as soon as symptoms show up. In some countries, effort was made to catch patients with mild symptoms, mainly by checking body temperature broadly in public places such as schools.

One of the key points is that, in order to discover an entire cluster of infections, the contact information from all individuals in the cluster is needed. For the most timely disease surveillance and prevention of its spread, all such contact information should be available at soon as a patient is confirmed. This suggests that, from the purpose of minimizing the disease spread, the best arrangement is to make all contact information available in real time, so that the process for discovering the entire cluster can take place immediately. In contrast, it is less than optimal if the contact information is requested one by one on a need basis as the tracing process progresses, which is the method of other tracing systems/proposals \cite{TraceTogether, AppleGoogle}.

A related key point is that what makes such tracing method powerful is the ability to discover an entire cluster, not just some members of the cluster. Under the assumption of universal adoption, there is no reason to discover anything less than the whole cluster. More fundamentally, it is the manifestation problem that leads to the need to trace an entire cluster because patients with mild or no symptoms may only be discoverable through their connections with other patients in the cluster who have more severe symptoms.

There is another view about the effectiveness of such tracing method, which is based on a consideration of the reproduction ratio. Once an infected person is identified, by downstream tracing, all the persons he/she infected can be immediately isolated from the general population. Thus, for all symptomatic cases, the reproduction ratio can effectively be reduced to zero. Asymptomatic patients can be identified by upstream tracing and testing. Then, by downstream tracing, the reproduction ratio for asymptomatic patients can also be reduced to zero. In Section \ref{sec:nearuniv}, we will use this view to argue for the effectiveness of the tracing method when the device adoption is not universal.

\subsection{Is Universal Adoption Necessary?} \label{sec:nearuniv}

In Section \ref{sec:universaladopt}, we argued that, under the assumption of universal adoption (i.e., that everyone carries a contact recorder), the proposed contact-tracing strategy will be completely effective in that it will be able to catch all growing clusters of infections. Since universal adoption is difficult to achieve and may be deemed draconian, we ask the natural question whether the strategy can remain effective if certain percentage of people are opt out, and if so, what that percentage is. Here, effectiveness is with respect to the original goal of keeping the number of infections low on a consistent basis. The quick answer is: Universal adoption is not necessary; however, to achieve the stated goal, the adoption rate needs to be very high. We will argue this formally. The often-used notations are summarized in Table \ref{tab:notations}.

\begin{table}[ht]
\caption{often-used notations}
\label{tab:notations}
\begin{center}
\begin{tabular}{|c|l|}
\hline
$R_0$ &  basic reproduction ratio \\ \hline
$\epsilon$ &  $\epsilon=1/R_0$ \\ \hline
$R_e$ &  {\em effective} reproduction ratio with tracing and quarantine \\ \hline
$N$ & a random variable representing the number of \\
& direct infections by a patient \\ \hline
$A, A_i$ & patient $A$ infects individuals $A_1, \ldots, A_N$ \\ \hline
$\chi$ & {\em effective} number of direct infections caused by a patient \\ \hline
$X_i$ & contribution ($0$ or $1$) to $\chi$ by $A_i$, and $\chi = \sum_{i=1}^N X_i$ \\ \hline
$\pi_0$ & probability that a cluster eventually stops growing \\
& under the condition of no mitigation measures \\ \hline
$\pi_1$ & $1-\pi_0$, the probability that a cluster grows to infinity \\ \hline
$p$ & adoption rate \\ \hline
$p^*$ & minimum required adoption rate such that $R_e < 1$ \\ \hline
$\nu$ & probability that a patient shows severe symptoms \\ \hline
$\mu_k$ & probability that at least one patient shows severe \\
& symptoms in a cluster of size $k$ \\ \hline
$EC(A_i)$ & eventual size of the cluster started by patient $A_i$ \\
& when there is no tracing \\ \hline
$Y$ & the random variable representing the cluster size \\
& when the first patient is tested \\ \hline
\end{tabular}
\end{center}
\end{table}

First, it is clear that the more people who carry contact recorders, the more effective the contact-tracing strategy is in curbing the spread of the virus. However, it may be the case that, at a certain adoption rate of the device, the tracing strategy can slow down the growth of the number of infections, but it is still an exponential growth, just a slow one. If the goal is to maintain a very low number of infections as a steady state, the effective reproduction ratio needs to be kept below 1. We will show that there is a cutoff value with respect to the required adoption rate. Below the cutoff value, the corresponding reproduction ratio is larger than 1; above it, the corresponding reproduction ratio is less than 1, in which case the number of infections will eventually go to zero. 

Suppose the adoption rate is denoted by $p$, which is the fraction of the population with contact recorders. Consider an arbitrary infected person $A$. Suppose $A$ goes on to directly infect $N$ other individuals, $A_1, \ldots, A_N$, where $N$ is a random variable. The expectation of $N$ is equal to $R_0$, where $R_0$ is the basic reproduction ratio. Under the proposed tracing strategy, we will define the {\em effective reproduction ratio} of $A$, denoted by $R_e$, which is equal to the expectation of $\chi$, the {\em effective} number of individuals directly infected by $A$. In general, $\chi$ will not be equal to $N$. If the cluster of infections started by $A_i$ can be fully discovered by tracing at some future time, the whole cluster, including $A_i$, will be isolated from the general population; in that case, $A_i$ does not contribute to $\chi$. The point is that, although $A$ has infected $A_i$, the effect of $A_i$ is entirely removed at some future time due to tracing and quarantine. It is as if $A_i$ does not exist when considering the spread of the disease.

More specifically, if $N=0$, we let $\chi=0$. Next, suppose $N \geq 1$. Let $X_i$ denote the contribution to $\chi$ by each $A_i$, and let $\chi = \sum_{i=1}^N X_i$. We let $X_i = 0$ if the cluster started by $A_i$ is fully traced (and removed) at some later time; we let $X_i = 1$, otherwise.

The $X_i$'s are IID, and they are independent of $N$. We then have $R_e = E[\chi] = E[N] E[X_i]=R_0 E[X_i] = R_0 P(X_i=1)$. We wish to have $R_e<1$ so that the spread of the disease will eventually stop.

Let $\pi_0$ be the probability that the cluster started by $A_i$ eventually stops growing {\em under the assumption that there are no mitigation measures of any sort}; let $\pi_1$ be the probability that the cluster grows to infinity under the same assumption. We have $\pi_0+\pi_1=1$.  For notational convenience, let $\epsilon = 1/R_0$.
The following theorem implies that our objective is achievable even without universal adoption.
\begin{theorem} \label{thm:noneunivadopt}
Suppose $\pi_1 > 1-\epsilon$. There exists $p_o < 1$ such that for all $p > p_o$, the effective reproduction ratio $R_e$ is less than 1.
\end{theorem}
\begin{IEEEproof}
Suppose $N \geq 1$ and $A$ infected individuals $A_i$, where $i = 1, \ldots, N$. 
Recall that $X_i=0$ if the cluster started by $A_i$ is fully traceable at some point. That happens if and only if (1) the cluster reaches some size $k$, (2) at least one patient in the cluster has caught the attention of the medical system and has been tested, and (3) everyone in the cluster has a contact recorder so that the whole cluster can be traced.

Let $EC(A_i)$ denote the eventual size of the cluster started by $A_i$ {\em if it is unimpeded}; that is, if there is no tracing and quarantine. $EC(A_i)$ can take a value from $[1, \infty]$. Let $Y$ be the random variable representing the cluster size when the first patient is tested. It is reasonable to think that, as long as the cluster grows indefinitely, some patient will eventually be tested. Therefore, we assume that, conditional on the cluster size grows to infinity, $P(Y < \infty) = 1$. We then have the following.
\begin{align}
& P(X_i=0) \nonumber \\
= & \sum_{k=1}^\infty P\{EC(A_i) = k \text{ and } Y \leq k \text{ and everyone } \nonumber \\
& \quad \quad \text{in the cluster of size $Y$ has a contact recorder} \} \nonumber \\
& + \sum_{k=1}^\infty P\{EC(A_i) = \infty \text{ and } Y = k \text{ and everyone } \nonumber \\
& \quad \quad \text{in the cluster of size $k$ has a contact recorder} \} \label{eq:pxieq0full} \\
\geq &  \sum_{k=1}^\infty P\{EC(A_i) = \infty \text{ and } Y = k \text{ and everyone } \nonumber \\
& \quad \quad \text{in the cluster of size $k$ has a contact recorder} \} \nonumber \\
= & \sum_{k=1}^\infty p^k P\{EC(A_i) = \infty \} P\{ Y=k | EC(A_i) = \infty \} \\
= & \pi_1 \sum_{k=1}^\infty p^k P\{ Y=k | EC(A_i) = \infty \}. \label{eq:2ndterm}
\end{align}
Now, by the monotone convergence theorem, $\sum_{k=1}^\infty p^k P\{ Y=k | EC(A_i) = \infty \} \to \sum_{k=1}^\infty P\{ Y=k | EC(A_i) = \infty \} = 1$ as $p \to 1$. Therefore, for any $\delta>0$, there exists $p_o < 1$ such that for all $p > p_o$, $P(X_i=0) > \pi_1 (1-\delta)$.

Now choose $\delta$ such that $\pi_1 (1-\delta) \geq 1-\epsilon$ and choose the corresponding $p_o$. We then have $P(X_i=0) > 1-\epsilon$ and therefore $R_e = R_0 P(X_i=1) < 1$ for all $p > p_o$.
\end{IEEEproof}

Theorem \ref{thm:noneunivadopt} implies that, the contact-tracing strategy can still achieve its goal of stopping the spread of the disease if some people, up to a certain fraction of the population, are opt out or otherwise unable to participate. Another implication is that if everyone participates but people occasionally do not carry the contact recorder on some of their outings -- again up to a certain threshold -- the tracing strategy will still work. In particular, exceptions can be made for some individuals or some occasions for which privacy protection is needed at the highest level, as long as they are made rarely (which we will discuss later).

For the parameter values of interest, $\pi_1$ is expected to be very close to $1$, and the condition of Theorem \ref{thm:noneunivadopt} is easily satisfied. As an example, suppose the number of individuals directly infected by a patient is a random variable with a Poisson distribution of mean $R_0$. It can be shown that $\pi_0$ is the solution of the equation $\log \pi_0 = R_0 (\pi_0 - 1)$ with a value less than 1 (see page 226 of \cite{Ross96}).\footnote{Throughout, $\log$ denotes the natural log.} Table \ref{tab:pi0pi1} shows the values of $\pi_0$ and $\pi_1$ for $R_0=3, 4, 5$, and $6$, which will be used later. We also see that $\pi_1$ is substantially greater than $1-\epsilon$ for these cases that are relevant to COVID-19.

\begin{table}[ht]
\caption{$\pi_0$ and $\pi_1$ under Poisson Distribution}
\label{tab:pi0pi1}
\begin{center}
\begin{tabular}{|c|c|c|c|}
\hline
$R_0$ & $\pi_0$ & $\pi_1$ & $1-\epsilon$ \\
\hline
3 &	0.0595  &	0.9405 &	0.6667 \\
4 &	0.0198  &	0.9802 &	0.75 \\
5 &	0.0070  &	0.9930 & 	0.8 \\
6 & 0.0025  & 	0.9975 &  	0.8333 \\
\hline
\end{tabular}
\end{center}
\end{table}

In the Poisson case, the condition of Theorem \ref{thm:noneunivadopt} is always satisfied due to the following result.
\begin{lemma}
Suppose the number of direct infections caused by a patient follows a Poisson distribution with mean $R_0$. Then, $\pi_0 < \epsilon$ and $\pi_1 > 1-\epsilon$. 
\end{lemma}
\begin{IEEEproof}
As we have always assumed, $R_0 > 1$ so that $0 < \epsilon < 1$. We know that $\pi_0 < 1$ and it satisfies $\log \pi_0 = R_0 (\pi_0 - 1)$. Consider the function $f(x) = \log x - R_0(x-1)$. We have $f(1)=0$, and $f(x) \to -\infty$ as $x \downarrow 0$. Now, $f'(x)=1/x - R_0$. Therefore, $f(x)$ increases on $(0, \epsilon)$ and decreases on $(\epsilon, 1]$. Moreover, $f(x)$ achieves the maximum on $(0,1]$ at $x=\epsilon$ and the maximum value is greater than $0$. Hence, there is exactly one solution to $f(x)=0$ on the interval $(0,1)$ and it lies on $(0, \epsilon)$. This solution is the required $\pi_0$.
\end{IEEEproof}

\subsection{Lower and Upper Bounds for Minimum Adoption Rate}

The next natural question is: How high need the adoption rate be in order to achieve the objective of stopping the spread of the disease? We will show that a very high adoption rate is needed, e.g., larger than $95\%$. Therefore, although there is some wiggle room that allows some people or some situations to be exempted, an appropriate public health message should encourage everyone to almost always carry the contact recorder when in public except on rare occasions when privacy is a must. 

Let $p^*$ denote the {\em minimum required adoption rate} such that $R_e < 1$, i.e., the smallest $p_o$ for Theorem \ref{thm:noneunivadopt}. We will show our answer to the question by calculating a lower bound and an upper bound for the minimum required adoption rate $p^*$. It turns out the two bounds are extremely close, and therefore, they are both good estimates of $p^*$.

The analysis requires a bit more assumptions than what is needed for Theorem \ref{thm:noneunivadopt}. Let $\nu$ be the probability that an infected person shows severe enough symptoms to be picked up by the medical system, where $0<\nu<1$. Let us assume that the patients are IID with respect to showing severe enough symptoms. When the size of a cluster of infected patients is equal to $k$, let $\mu_k$ denote the probability that at least one patient shows severe enough symptoms (who will be tested). Then, by the definition of $\nu$ and the IID assumption,
\begin{align}
\mu_k = 1-(1-\nu)^k.
\end{align}

\subsubsection{Lower Bound}

We will start from (\ref{eq:pxieq0full}). The second term on the right hand side (\ref{eq:pxieq0full}) can be written as the right hand side of (\ref{eq:2ndterm}). Conditional on that the cluster size grows to infinity when unimpeded, we have $P(Y=k) = (1-\nu)^{k-1} \nu$. Continuing from (\ref{eq:2ndterm}), we get
\begin{align}
& \ \pi_1 \sum_{k=1}^\infty p^k P\{ Y=k | EC(A_i) = \infty \} \nonumber \\
= & \ \pi_1 \sum_{k=1}^\infty p^k (1-\nu)^{k-1} \nu \nonumber \\
= & \ \pi_1 \frac{\nu p}{1 - p(1-\nu)}. \label{eq:infres}
\end{align}

We next work on the first term on the right hand side of (\ref{eq:pxieq0full}), which can be written as
\begin{align}
& \sum_{k=1}^\infty P\{EC(A_i) = k \text{ and } Y \leq k \text{ and everyone } \nonumber \\
& \quad \quad \text{in the cluster of size $Y$ has a contact recorder} \} \nonumber \\
= & \sum_{k=1}^\infty \sum_{j=1}^k p^j P\{EC(A_i) = k, Y = j \} \nonumber \\
= & \sum_{k=1}^\infty \sum_{j=1}^k p^j P\{EC(A_i) = k\} P\{ Y = j | EC(A_i) = k \} \nonumber \\
= & \sum_{k=1}^\infty \sum_{j=1}^k p^j P\{EC(A_i) = k\} \frac{(1-\nu)^{j-1} \nu}{\mu_k} \nonumber \\
= & \sum_{k=1}^\infty \frac{1-(p(1-\nu))^k}{1-(1-\nu)^k} P\{EC(A_i) = k\}  \frac{\nu p }{1 - p(1-\nu)}. \label{eq:finitebd}
\end{align}

\begin{lemma} \label{lem:maxratio}
For $k \geq 1$, $\frac{1-(p(1-\nu))^k}{1-(1-\nu)^k}$ is decreasing in $k$. It is maximized at $k=1$ and the maximum is equal to $(1-p(1-\nu))/\nu$.
\end{lemma}
\begin{IEEEproof}
We will tentatively view $k$ as a real-valued variable and take the derivative with respect to $k$. 
\begin{align}
& \Bigl( \frac{1-(p(1-\nu))^k}{1-(1-\nu)^k} \Bigr)' \nonumber \\
= & \frac{-p^k(1-\nu)^k (1-(1-\nu)^k) \log(p(1-\nu))}{(1-(1-\nu)^k)^2} \nonumber \\
& + \frac{(1-p^k(1-\nu)^k) (1-\nu)^k \log(1-\nu) }{(1-(1-\nu)^k)^2}.
\end{align}
To show the above derivative is less than $0$ for $k \geq 1$, it is enough to show
\begin{align}
& -p^k (1-(1-\nu)^k) \log(p(1-\nu)) \nonumber \\
& + (1-p^k(1-\nu)^k) \log(1-\nu)  < 0, \nonumber
\end{align}
which simplifies to
\begin{align}
& -p^k \log(p(1-\nu)) + p^k  (1-\nu)^k \log p + \log(1-\nu)  < 0. \label{eq:simpcompk}
\end{align}
At $k=0$, the left hand side of (\ref{eq:simpcompk}) is equal to $0$. The derivative of the left hand side is equal to $-p^k \log p \log(p(1-\nu)) + p^k  (1-\nu)^k \log p \log(p(1-\nu))< 0$. Hence, the left hand is decreasing in $k$ and its value is less than $0$ for all $k > 0$.
\end{IEEEproof}

Using the result of Lemma \ref{lem:maxratio}, the right hand side of (\ref{eq:finitebd}) is less than or equal to
\begin{align}
& \max_{k \geq 1} \frac{1-(p(1-\nu))^k}{1-(1-\nu)^k} \sum_{k=1}^\infty P\{EC(A_i) = k\}  \frac{\nu p }{1 - p(1-\nu)} \nonumber \\
= & p \pi_0. \label{eq:finbd}
\end{align}

Then, from (\ref{eq:pxieq0full}), (\ref{eq:finbd}) and (\ref{eq:infres}),
\begin{align}
& P(X_i=0) \nonumber \\
\leq & p \pi_0 + \pi_1 \sum_{k=1}^\infty p^k P\{ Y=k | EC(A_i) = \infty \} \nonumber \\
\leq & p \pi_0 + \pi_1 \frac{\nu p}{1 - p(1-\nu)}.
\end{align}
To have $R_e = R_0 P(X_i=1) < 1$, we need $P(X_i=0) > 1-1/R_0 = 1-\epsilon$, which implies
\begin{align}
p \pi_0 + \pi_1 \frac{\nu p}{1 - p(1-\nu)} > 1-\epsilon. \label{eq:lbinequa}
\end{align} 
The inequality (\ref{eq:lbinequa}) can be rewritten as follows, involving a quadratic function of $p$.
\begin{align}
\pi_0 (1-\nu) p^2 - (\pi_0 + \pi_1 \nu + (1-\epsilon)(1-\nu))p + (1-\epsilon) < 0.
\end{align}
The value of the quadratic function at $p=1$ is equal to $-\epsilon \nu < 0$. Therefore, one of the solutions of the quadratic function is greater than 1 and the other is less than 1. The latter solution will be denoted by $\underline{p}$. Then, (\ref{eq:lbinequa}) holds if $\underline{p} < p \leq 1$. We can compute $\underline{p}$ using the standard expression for solving a quadratic equation, which gives
\begin{align}
& \underline{p} = \frac{1}{2 \pi_0 (1-\nu)} \Bigl( \pi_0 + \pi_1 \nu + (1-\epsilon)(1-\nu) \nonumber \\
& - \sqrt{(\pi_0 + \pi_1 \nu + (1-\epsilon)(1-\nu))^2-4\pi_0 (1-\nu)(1-\epsilon)} \Bigr). \label{eq:plbd}
\end{align}
The condition $p > \underline{p}$ is necessary for having $R_e < 1$. Therefore, $\underline{p}$ is a lower bound for the minimum required adoption rate, i.e., $p^* \geq \underline{p}$.

\subsubsection{Upper Bound}

We will start with (\ref{eq:finitebd}). Using the result of Lemma \ref{lem:maxratio}, for any $k \geq 1$, $\frac{1-(p(1-\nu))^k}{1-(1-\nu)^k}$ is greater than or equal to the limiting value, which is equal to $1$, as $k \to \infty$. Therefore, the right hand side of (\ref{eq:finitebd}) is greater than or equal to
\begin{align}
\sum_{k=1}^\infty P\{EC(A_i) = k\}  \frac{\nu p }{1 - p(1-\nu)} = \pi_0 \frac{\nu p }{1 - p(1-\nu)}. \label{eq:finlbd}
\end{align}

Then, from (\ref{eq:pxieq0full}), (\ref{eq:infres}) and (\ref{eq:finlbd}),
\begin{align}
P(X_i=0) \geq  \frac{\nu p}{1 - p(1-\nu)}.
\end{align}
For any $p$ value that satisfies $\frac{\nu p}{1 - p(1-\nu)} > 1 - \epsilon$, we have $R_e<1$. This gives  
\begin{align}
p > \bar{p} \triangleq \frac{1}{1-\nu + \frac{1}{1-\epsilon} \nu}=\frac{1}{1+\frac{\epsilon}{1-\epsilon} \nu}. \label{eq:pubd}
\end{align}
Note that $\bar{p}$ defined in (\ref{eq:pubd}) is the $p_o$ for Theorem \ref{thm:noneunivadopt} and it is always less than 1. We see that $\bar{p}$ is an upper bound of the minimum required adoption rate in order to achieve $R_e<1$, i.e., $p^* \leq \bar{p}$.

\subsubsection{Numerical Results}

For the numerical results, we assume the number of direct infections by an arbitrary patient, i.e., the random variable $N$ in Section \ref{sec:nearuniv}, follows a Poisson distribution with mean $R_0$. Table \ref{tab:numreslbub} shows sample values for the lower and upper bounds computed based on (\ref{eq:plbd}) and (\ref{eq:pubd}), respectively. We see that the two bounds are very close under the chosen set of parameters. Therefore, we have a pair of very good estimates for the true value of the minimum required adoption rate $p^*$.

\noindent  {\bf Remark.} The Poisson assumption is not required to derive the general expressions for $\underline{p}$ and $\bar{p}$ in (\ref{eq:plbd}) and (\ref{eq:pubd}), respectively. For the numerical results, we need to calculate the values of $\pi_0$ and $\pi_1$ used in the expression for $\underline{p}$, and the Poisson assumption is needed for such calculation. Since the expression for $\bar{p}$ does contain $\pi_0$ and $\pi_1$, the numerical values for $\bar{p}$ are also the upper bound for general distributions. Therefore, even when we do not know the distribution for the number of direct infections, $\bar{p}$ can still be used to provide a useful upper bound.

\begin{figure}[ht]
\centering
\includegraphics[width=3in]{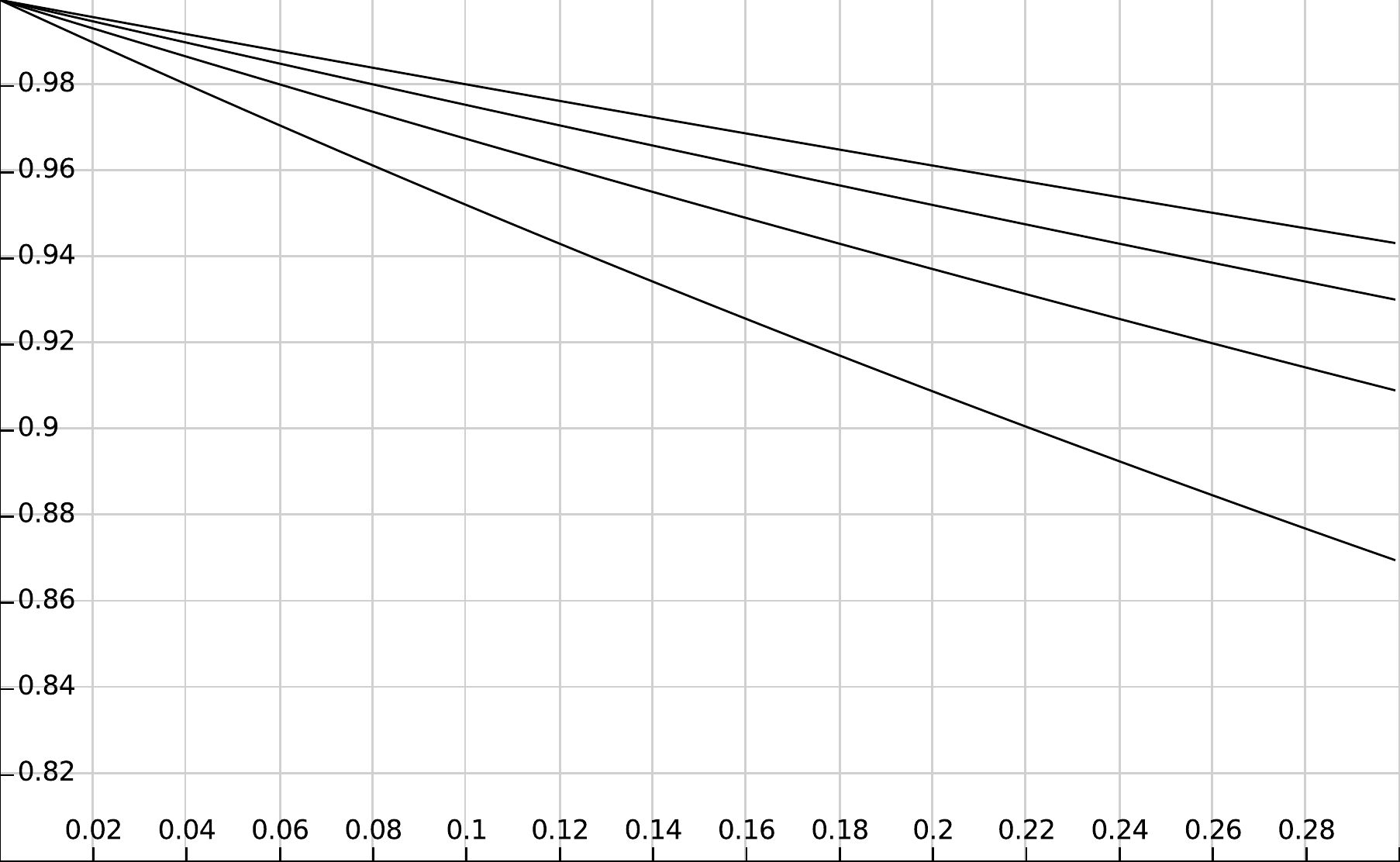}
\caption{The upper bound $\bar{p}$ versus $\nu$. The four curves correspond $R_0=3, 4, 5$ and $6$, respectively, in the bottom-to-top order.}
\label{fig:pubvsnu}
\end{figure}

The set of parameters used to generate the numerical results is relevant to COVID-19. Table \ref{tab:numreslbub} shows the cases for $R_0=3, 4, 5$ and $6$. These values should be enough to cover the range of estimates for the true $R_0$ of COVID-19. For COVID-19, if $50\%$ of the infected individuals are asymptomatic and $10\%$ of the symptomatic patients develop severe conditions, then, on average, one out of a cluster of 20 patients has severe symptoms, which can be picked up and tested by the medical system. In that case, $\nu$ can be taken as $0.05$. Table \ref{tab:numreslbub} also shows the result for $\nu=0.02$, which is relevant if the true asymptomatic cases are drastically more than the current estimate or if the medical system is overly relaxed and not picking up infection cases frequently enough. The result for $\nu=0.1$ shows the benefits if the disease surveillance system can pick up patients with mild symptoms quickly, for instance, by widespread temperature measurement. The required adoption rate can be lowered to the more manageable $95\%$ range. To pursue the benefit further, in  Fig. \ref{fig:pubvsnu}, we plot the upper bound $\bar{p}$ against a wider range of $\nu$. It shows that, if we are lucky enough that the true $R_0$ is around 3 and if we are vigilant enough to catch infection cases early, the required adoption rate can be brought to the $90\%$ range.

\begin{table}[ht]
\caption{Numerical Results for the Lower and Upper Bounds}
\label{tab:numreslbub}
\begin{center}
\begin{tabular}{|c|c|c|c|}
\hline
$\nu$ & \ $\epsilon$ \ \ ($R_0$) & Lower Bound $\underline{p}$ & Upper Bound $\bar{p}$ \\
\hline
0.1 & 	1/3 \ (3) &	0.94865	&	0.95238 \\
	&	1/4 \ (4) &	0.96701	&	0.96774 \\
	&	1/5 \ (5) &	0.97543	&	0.97561 \\
	& 	1/6 \ (6) &	0.98034	& 	0.98039 \\
\hline
0.05 &	1/3 \ (3) &	0.97347 &	0.97561 \\
	 &	1/4 \ (4) &	0.98320	&	0.98361 \\
	 &	1/5 \ (5) &	0.98755	&	0.98765 \\
	 & 	1/6 \ (6) &	0.99007	& 	0.99010 \\
\hline
0.02 & 	1/3 \ (3) &	0.98917	& 	0.99010 \\
	 & 	1/4 \ (4) &	0.99320	& 	0.99338 \\
	 & 	1/5 \ (5) &	0.99498	& 	0.99502 \\
	 & 	1/6 \ (6) &	0.99600	& 	0.99602 \\
\hline
\end{tabular}
\end{center}
\end{table}

We next explain why the required adoption rate is usually high. Take the example of $\nu=0.05$, which corresponds to the case where, on average, there is one patient with severe symptoms in a cluster of 20 infected cases. One expects that the cluster size has to grow to about $20$ individuals before a severe case shows up with a reasonable chance. To have a high chance to observe a severe case, the cluster size needs to be larger than the average number of $20$, such as $40$. Once a severe case is found, we want to be able to trace the entire cluster with a high enough probability. For instance, at a probability of $0.75$, which is needed when $R_0=4$, the required $p$ for $p^{40} > 0.75$ is $0.993$.

\subsection{What happens if the adoption rate is less than $p^*$?}

It is important to emphasize that when the adoption rate $p$ is sufficiently high, i.e., $p > p^*$, the effective infection ratio $R_e$ is less than 1. Even without any additional disease mitigation measures (other than testing and quarantine) such as social distancing, the spread of the disease will eventually be stopped and the number of infection cases will go down to zero or be maintained at the near-zero level indefinitely. In that sense, the society can truly return to normal.

When the adoption rate is less than $p^*$, the disease will be able to spread exponentially in the absence of additional measures. However, the proposed tracing strategy will still be useful even at a moderate adoption level. It solves some of the problems encountered in manual tracing, by making some part of the tracing more automatic, more timely, and less labor intensive. The higher the adoption rate of contact recorders, the less effort is needed in manual tracing. With the help of contact recorders, it is still possible to bring the disease under control, i.e., to achieve $R_e<1$; but other measures such as social distancing need to be retained as part of the solution. Nevertheless, it is hopeful that the device-based tracing can help so much that other measures can be reduced in intensity or frequency of application. The society can return to partial normalcy instead of full normalcy.

\section{Interim and Long-Term Solutions}
\label{sec:interimlongterm}

\subsection{Some Features That the Contact Recorder Should Have}

It is not automatic that the contact recorder should be implemented as a phone app. We will first go over a list of requirements or desirable features such a device should have.

\begin{itemize}

\item	The device should record the contacted ID, the starting time of that contact, the duration of the contact, and the distance of the contact. It only needs to be able to measure distance within a short range, i.e., under several meters. 

\item	Although the device mainly relies on a kind of proximity sensor for short-range contact detection and distance measurement, it should also have the GPS capability to record GPS trajectories. Although the GPS data is not useful for detecting short-range contact, it is useful in many other ways. For instance, with the help of mapping information, the server can deduce whether a contact occurs in a building, and whether the building is crowded at that time. If the contact occurs in a crowded building, there may be other means of virus transmissions besides person-to-person contact, such as touching shared surfaces. A decision can be made whether to send warning or quarantine directives more broadly, or to seal off the building.

\item	The device should be able to communicate with the central server in real time. In particular, the contact information should be available to the server in real time so that the server can conduct sophisticated tracing and do so as broadly as needed. 

\item	The communication should be two-way so that the device is able to receive information or download data from the server.

\item	The device should be inexpensive, small, light and energy efficient so that it is not too much of a burden to acquire, wear and maintain it.

\item	Preferably, the device is able to monitor body temperature and other vital signs.

\item	The device and the whole system should have necessary security and privacy-protection mechanisms.

\item	The software of the device should be upgradable automatically.
\end{itemize}

\subsection{Interim Solution: Mobile Phones as Contact Recorders}

Most of the newer phones are equipped with Bluetooth Low Energy (BLE), which allows them to sense the presence of each other when they come into contact, which is a form of proximity sensing. Such phones satisfy most of the requirements listed earlier. However, one of the main problems is that their distance measurement is not accurate because the distance is estimated solely based on RSSI (received signal strength indicator). As an example, in the current implementation of iOS's iBeacon, the sensed distance is categorized into immediate (i.e., the two phones are pretty much next to each other), near (maybe up to a few meters), and far (anything above near). To act as contact recorders, ÔimmediateÕ and ÔnearÕ are close to the intended sensing range. If two phones fall into the immediate range, it is almost certainly the case that the two individuals are in close contact. If they fall in the near range, the individuals may be under a meter to a few meters apart. They may or may not be in the infection range. In those cases, timing information such as the duration of contact can be combined with the distance information when the server assesses the risk of infection. For instance, if the contact lasted 30 minutes in the near range, then the risk of infection may be considered high. Overall, the contact-tracing strategy can still be effective if the inaccuracy in distance measurement is compensated by more aggressive risk evaluation, testing, and quarantine. 

The phone-based system has many advantages. On the phone side, the system is easy and quick to deploy as people only need to download and activate a phone app. The app is relatively easy to program, when compared with building a new system from scratch. It is relatively easy to add sophisticated features, including strong security and privacy protection. The app is readily upgradable so that the capabilities of the system can be gradually improved. It costs nothing for people who already have a phone with the right generation of Bluetooth technologies. The phone app can communicate in real time with a central server or servers run by either an authority or a private company on behalf of the authority. Data collection, analysis and decision-making can be done in real time. 

\subsection{Enhancement: Adding Fixed-Location Beacons}
\label{sec:beacon}

The direct mobile-to-mobile contact detection can be supplemented by beacon-to-mobile communication for localization. The BLE-based beacons are one of the upcoming technologies used in proximity-sensing systems, because they interact with mobile phones very well. In fact, the same technology can be used for a mobile phone to detect either another mobile phone or a beacon. In addition, these beacons are inexpensive, small, easy to deploy, and they can last up to several years with a single battery (for instance, see products of Estimote \cite{Estimote}). Their main function is to broadcast beacon signals periodically, e.g., once every second. A smart phone can detect the beacon signal when it comes into a sufficiently close range.

Beacons can be installed widely in major public venues, especially in closed spaces such as different parts of a conference hall or a theater, different subway cars, or buses. Once a mobile phone detects a nearby beacon, it records the beacon ID and time of encounter. In the backend, the server has information on where exactly that beacon is located. When a patient is discovered, the beacon IDs recorded on his/her phone can be used to determine the places that the patient has been to and the time of the visits. For instance, with beacons installed on subway trains, it is possible to identify which car of a subway train the patient has traveled on. The subway car may be disinfected if needed. Moreover, if the virus can be transmitted over a longer distance in aerosols or in droplets carried by air flows from air conditioner systems, which is likely the case for the coronavirus, people who are in the same closed space but not within the close-range contact distance are also at risk. Such beacon-based tracing can be used to discover those individuals.

The beacon-based scheme is similar to the QR-code-based scheme used in countries such as China and South Korea. When entering a public place such as a building or a bus, each individual uses a phone app to scan a QR code displayed at that location. This way, the backend server knows who entered the place at different time. If an individual is later confirmed to be infected, the server is able to identify all the people who were together with the infected individual at a location. Although the QR codes are easier to deploy, the beacon-based scheme has the advantage that the phones automatically scan beacon signals. Manual scanning of QR codes can become tiresome if an individual has to scan many times a day at different places. Without service personnel stationed at each QR-code location, people may eventually skip scanning the code. The beacon-based scheme is far more scalable, making it possible to cover the public space finely.

\subsection{Longer-Term Solutions}

Even after the current crisis, other pandemics will strike again. Vaccine development, effective treatment and even testing will always be lagging for a new infectious disease. However, quarantine and contact tracing are infection-control measures that can be applied immediately. The mobile device and the tracing strategy outlined here may become part of the public health measures to combat future pandemics. At any moment, the society should have a version of the mobile device and a concrete tracing strategy that are readily deployable. Meanwhile and when time allows, the design of the mobile device can be continuously refined. In particular, there is a need to improve the accuracy of distance sensing, down to centimeter range, possibly by using different families of sensors or with the help of localization techniques such as triangulation. The inclusion of more biosensors can be very useful for disease surveillance and patient monitoring.

In the long run, the mobile device may or may not be part of a mobile phone. It could be a standalone device that communicates with a phone app, or it could be a completely independent device communicating with the central server directly via cellular networks. A standalone device has some advantages over the phones. It can be made smaller, and therefore, more wearable, like a watch, a name card or a pin. It may be simpler to use so that people who are unfamiliar with smart phone operations can participate. It may be cheaper than a phone. All these properties make them suitable for children of young ages. A standalone device can also be more visible when a person wears it. Such visibility is important for the intended use. In all likelihood, standalone devices may coexist with phone-based devices.

\subsection{Security and Privacy}

Security and privacy experts have quickly entered the conversations about proximity-based tracing through phone apps, and they are offering their thoughts and preliminary solutions for data security and privacy preservation \cite{CIY20, DPPT, Kuhn2020CovidNT, Bell2020TraceSecureTP}. Privacy preservation is also central in the proposed APIs by Apple and Google to support contact-tracing apps \cite{AppleGoogle}. In this paper, we will not review the details of what have been considered, but refer the readers to the relevant writings. We would like to emphasize that privacy preservation depends on the assumed operations of the tracing system. For instance, the APIs of Apple and Google address how to preserve anonymity between the two parties of a contact, and how to ensure consent before contact data is uploaded to the server. Simply put, the privacy and security features will depend on what data different components of the system collects, the movement of the data, and how the data are used. In light of the stated goal of this paper, we envision a system with maximum liberty in data collection and use. The server at any time should have all the needed information to trace and identify a cluster of contacts, and know how to reach them for testing. The questions are, under that setting, to what extent privacy can be preserved and how.

Although the details are complex, some basic principles seem to be uncontroversial. The authority must promise that the collected contact information will only be used for infectious disease control. It will not be used in any way that leads to any harm to the individuals. The data will be deleted or anonymized after they are no longer useful for disease control, e.g., after several weeks to several months.\footnote{For instance, in Singapore's TraceTogether, the contact information of an infected individual is uploaded to the government server only after consent, is encrypted and kept for up to 21 days.} The data must be secure against theft through proper encryption. How to enforce these principles is both a technological issue and a legal issue.

For individuals who are not comfortable with having their contact information kept on the server for any amount of time or in any form, the system can still be useful while guaranteeing full protection of privacy. For instance, the server can store non-identifiable information of consenting patients, e.g., hashed values of their device IDs. The individuals who do not wish to upload their contact information can download the infection information from the server and check whether they have been in contact with any infected person. The APIs by Apple and Google highlight this particular approach. However, in order to bring the reproduction ratio below 1, such individuals must constitute a very small fraction of the population. 

\section{Conclusions and Discussion}
\label{sec:conclude}

This paper presents a best-case scenario about what proximity-sensing-based contact tracing can accomplish. It can achieve way more than what casual thinking may suggest; but the requirement is a high adoption rate. In particular, with a high enough adoption rate, there is likely no need for social distancing. One of our main purposes is to raise the aspiration level when applying such a method of tracing. All relevant parties of the society should put more effort into adopting the method and reaching a high adoption rate.

Although advocating for universal adoption may seem drastic and the target may seem difficult to reach, upon some thinking, the situation is not so different from other universal public-health measures, such as vaccination or mask-wearing in public. The latter has already happened in some cities and countries in the West. It is likely that universal adoption, or at least encouragement of doing so, can happen in some cities, states or even countries.

It is possible that the minimum required adoption rate can be lower than what is reported in this paper. If so, that could mean the tracing strategy can be fully effective at a more easily achievable adoption rate. The readers are invited to find opportunity to lower the bound. One likely place to look into is whether the definition of $R_e$ is the most relevant. 

Given the main theme of the paper, we would like to suggest privacy and security experts work more under the assumptions of universal adoption, real-time upload of contact information and automated detection of infection clusters. For tracing clusters of infections as quickly as possible, it will not work well if the contact information of some individuals stays on their phones and is uploaded only after consent. The aim should be to develop mechanisms for ensuring privacy under these new assumptions and for ensuring that the system is used only for the intended purpose and intended time frame (21 days of user data retention, etc.). 

Although the contact recorders help to solve very difficult problems and save a great deal of effort in contact tracing, success also depends on other potentially costly operations. In particular, the total number of contacts made by an infection cluster is likely many times more than the contacts that lead to actual infections. Testing must be able to keep up with the large number of contacts in order to correctly identify actual infections. Good quarantine operation is another costly component. Furthermore, to be able to detect infection clusters before they become too large, disease surveillance needs to be strengthened, for instance, by widespread temperature checking. To cope with the cost, much can be done to improve the tracing system as operational experiences accumulate. For instance, the system should adapt and learn to become more accurate in assessing infection risk, such as what sort of contacts are more likely to lead to infection, so as to reduce the number of tests needed.

In practice, there are various constraints and concerns that may limit the adoption rate to be below the minimum required rate. Even so, the device-based tracing approach will still be very useful. In situations where it is applicable, it is a great improvement over manual tracing. Depending on the actual adoption rate, we may not be able to end social distancing, but we may be able to greatly relax it. It is possible that a combination of device-based tracing, manual tracing, and some mild-form social distancing together can bring back much normalcy. As an evidence, even without device-based tracing, some Asian regions appear to have brought the reproduction ratio very close to 1 with intensive manual tracing, strong disease surveillance, and mild social-distancing measures. If device-based tracing is adopted, it will not only save much of the effort of manual tracing, but it may be able to bring the reproduction ratio decisively below 1 and further reduce the need for social distancing. The conjecture remains to be proven. Simulation studies can provide some early indication.

Throughout, we have mostly considered disease spreading through person-to-person direct contact. COVID-19 can also spread through other means. In Section \ref{sec:beacon}, we discussed using beacons to deal with spread over longer distance in closed space, such as some form of airborne transmissions in restaurants or conference halls. The tracing for those situations works similarly as that for close-distance human contacts, except that the number of individuals that need to be screened is likely much larger. There are also transmissions through shared surfaces, such as shared coffee tables or elevator buttons. The beacon approach can also be useful for some of those situations to pinpoint the locations of transmissions. There will be situations where none of the device-based approaches can catch the origin of infection. In our framework, we could view each of such occasions as the start of a cluster. The first patient of the cluster will later be identified through tracing after transmissions through person-to-person contact begin. In the end, it is still the person-to-person direct transmissions that are of the primary concern, because both sides of each transmission are mobile. Furthermore, if transmissions of this type can be dealt with effectively, fewer people will be infected and less of the shared environment will be contaminated.

\bibliographystyle{IEEEtran}
\bibliography{contact}

\end{document}